\documentclass[conference]{IEEEtran}
\IEEEoverridecommandlockouts
\usepackage{amsmath,amsfonts}
\usepackage{graphicx}
\usepackage{textcomp}
\usepackage{xcolor}
\usepackage{color, colortbl}

\usepackage[utf8]{inputenc}
\usepackage{verbatimbox}
\usepackage{multirow}
\usepackage{makecell}
\usepackage{subcaption,caption}
\usepackage{adjustbox}
\usepackage[scientific-notation=true]{siunitx}
\usepackage{todonotes}
\usepackage{balance}
\usepackage{cite}
\usepackage{amsmath,amssymb,amsfonts}
\usepackage[Algorithm,ruled]{algorithm}
\usepackage{url}
\usepackage[noend]{algpseudocode}
\usepackage{graphicx}
\usepackage{textcomp}
\usepackage{xcolor}
\usepackage{booktabs}
\usepackage{xspace}
\usepackage{enumitem}
\usepackage{listings}
\usepackage{pifont}
\usepackage{fancyvrb}
\usepackage{framed}
\usepackage{xcolor}

\definecolor{codered}{HTML}{D9534F}
\definecolor{codegreen}{HTML}{5CB85C}
\definecolor{codeblue}{HTML}{0000FF}

\def\BibTeX{{\rm B\kern-.05em{\sc i\kern-.025em b}\kern-.08em
    T\kern-.1667em\lower.7ex\hbox{E}\kern-.125emX}}
    
\definecolor{m1}{RGB}{224, 250, 255}
\definecolor{m2}{RGB}{219, 224, 255}
\definecolor{m3}{RGB}{252, 255, 220}
\definecolor{m4}{RGB}{249, 214, 177}

\newtheorem{definition}{Definition}

\begin{document}

\title{Rare-Seed Generation for Fuzzing
\thanks{This material is based on research sponsored by NSF under grants CCF-2008660, CCF-1901098 and CCF-1817242. The U.S. Government is authorized to reproduce and distribute reprints for Governmental purposes notwithstanding any copyright notation thereon. The views and conclusions contained herein are those of the authors and should not be interpreted as necessarily representing the official policies or endorsements, either expressed or implied, of the U.S. Government.}
}

\author{\IEEEauthorblockN{Seemanta Saha, Laboni Sarker, Md Shafiuzzaman, Chaofan Shou, Albert Li, Ganesh Sankaran, Tevfik Bultan}
\IEEEauthorblockA{\textit{Department of Computer Science} \\
\textit{University of California, Santa Barbara}\\
Santa Barbara, CA, US \\
\{seemantasaha,labonisarker, mdshafiuzzaman, shou, albert\_li, ganesh, bultan\}@ucsb.edu}
}

\maketitle

\begin{abstract}
Starting with a random initial seed, fuzzers search for inputs that trigger bugs or vulnerabilities. 
However, fuzzers often fail to generate inputs for program paths guarded by restrictive branch 
conditions. In this paper, we show that by first identifying rare-paths in programs (i.e., program paths with path constraints that are unlikely to be satisfied by random input generation), and then, generating inputs/seeds that trigger rare-paths, one can improve the coverage of fuzzing tools. 
In particular, we present techniques 1) that identify rare paths  
using quantitative symbolic analysis,
and 2) generate inputs that can explore these rare paths using path-guided concolic execution.
We provide these inputs as initial seed sets to three state of the art fuzzers. Our experimental evaluation on a set of programs (that contain a lot of restrictive branch conditions) shows that the fuzzers achieve better coverage with the rare-path based seed set compared to a random initial seed.

\end{abstract}


\section{Introduction}



Testing software in order to assure its dependability and security is one of the most fundamental problems in software engineering. Fuzz testing has emerged as one of the effective testing techniques for achieving code coverage and finding bugs and vulnerabilities in software. 
Unfortunately, existing fuzzers often fail to generate inputs for program paths guarded by restrictive branch conditions.
To pass through branch conditions most greybox fuzzers~\cite{afl, afl-fast, redqueen, fairfuzz, vuzzer, mopt} focus on input mutation strategies. On the other hand, hybrid fuzzers~\cite{driller,digfuzz} switch to symbolic execution in order to solve path constraints when fuzzing gets stuck.  Identifying the likelihood of the fuzzer getting stuck is a crucial problem for hybrid approaches, and using the fuzzer itself for this purpose (by monitoring fuzzing behavior) requires a lot of time to explore deeper paths.







Both input mutation-based fuzzers~\cite{fairfuzz} and hybrid fuzzers~\cite{digfuzz} focus on identifying rare paths in the program  to explore. They either use mutation strategies~\cite{fairfuzz} or symbolic execution~\cite{digfuzz} to generate inputs that can explore the rare paths. Both of these techniques identify rare paths based on the inputs generated and branches covered during fuzzing (for example, AFL~\cite{afl}). 
Note that, it may take a long time to generate a value that triggers a branch if the branch condition is very restrictive. It is difficult to separate infeasible paths from feasible but rare paths via input mutation. 

In this paper, we propose a lightweight whitebox analysis to identify rare paths in programs and then guide symbolic execution to generate inputs to explore these rare paths. Our approach avoids the shortcomings of mutation-based greybox fuzzers and hybrid fuzzers by generating inputs for rare paths beforehand, and it avoids the shortcomings of whitebox fuzzers based on symbolic execution by reducing the cost of symbolic analysis.

We present a heuristic for identifying rare paths where we use control flow analysis, dependency analysis and model counting on branch constraints to transform a control flow graph to a probabilistic control flow graph. Then, we compute path probabilities by traversing the probabilistic control flow graph and identify the rare (low-probability) paths. 


To improve the rare path analysis we introduce a new type of control flow paths (which we call II-paths) which is a combination of intra- and inter-procedural program paths, providing a balance between breadth first and depth first traversal of program paths. 

We guide concolic execution using rare paths to generate inputs that trigger these rare behaviors. As the last step of our approach, we provide the set of inputs from our analysis as the initial seed set to a fuzzer. This enables the fuzzer to explore the rare paths immediately, resulting in better coverage compared to randomly generated initial seed sets. 
Our approach can be integrated with all existing fuzzers that rely on initial seeds. 




Our contributions in this paper are as follows:
\begin{itemize}
    \item A new technique for identifying rare paths in programs using a lightweight quantitative symbolic analysis.
    \item A new type of control flow paths (II-paths) to improve efficiency and effectiveness of the rare path analysis.
    \item Algorithms for path-guided concolic execution.
    \item Rare-path guided fuzzing approach where the initial seed set for a given fuzzer is generated with rare-path analysis.
    \item Experimental evaluation of the proposed techniques on existing fuzzers AFL++, FairFuzz and DigFuzz, demonstrating coverage improvement achieved by the proposed rare-path guided fuzzing approach.
\end{itemize}

Rest of the paper is organized as follows. In section~\ref{sec:overview}, we provide overview of our technique. We explain program path analysis, heuristic to identify rare paths and input generation using the rare paths on section~\ref{section:program-paths},~\ref{section:rare-paths} and~\ref{section:seed-generation} respectively. We discuss our implementation and experimental evaluations on section~\ref{sec:implmentation} and~\ref{sec:experiments} respectively. In section~\ref{sec:relwork} we present the related works and finally conclude in section~\ref{sec:conclusion}.


\section{Overview}
\label{sec:overview}



Consider the running example in Fig.~\ref{fig:running-example} which is a shortened version of a code structure found in \texttt{libxml2}. The \texttt{main} procedure of the program reads a string as an argument. It checks if the first 3 characters of the string is \texttt{DOC} or not. If the first 3 characters of the string is \texttt{DOC}, it parses the string starting from the 4th character. First, it goes inside the \texttt{parse\_cmt} procedure and it checks if the 4th character is \texttt{<} or \texttt{>} and skips if it is. Then, the program comes back to the \texttt{main} procedure and goes inside the \texttt{parse\_att} procedure. In the \texttt{parse\_att} procedure, the program looks for the character sequence \texttt{ATT}. If it finds this sequence, it goes deeper into the program and executes more functionalities. To summarize, the program is trying to find two specific sequences of characters: first \texttt{DOC} and then \texttt{ATT} and if it can find these two sequences, it can execute more functionalities.


A mutation based fuzzer, such as AFL, starting with a random initial seed will require a lot of mutations to get to an input containing sequences \texttt{DOC} and \texttt{ATT}. We run AFL 5 times on the running example for an hour. 4 out of 5 times, AFL cannot generate an input containing sequences \texttt{DOC} and \texttt{ATT}. AFL can generate inputs such as \texttt{DOC}, \texttt{DOC\textless}, \texttt{DOC>}, \texttt{DOCA} etc. Though coverage guided mutation helps to reach these inputs, AFL can not generate the desired sequences as it mutates randomly and breaks already found sequences to inputs likes \texttt{DAC} and \texttt{DOCQ} etc.




Now, let us explain how rare path analysis can guide a mutation-based fuzzer to achieve more coverage given a time budget. To perform rare path analysis on the running example program, we first extract the control flow graph and then we collect control flow paths of the program. At this point, we can use two well known existing techniques for control flow analysis to collect paths: intra-procedural control flow analysis and inter-procedural control flow analysis. 
Control flow graphs for the code in Fig.~\ref{fig:running-example} are shown in Fig.~\ref{fig:interprocedural-cfg}.

First, we collect paths using intra-procedural control flow analysis (paths from 1 to 5 in Table~\ref{tab:intra-inter-paths}). Among these paths, we find that path 4 is the rarest one. We identify rarity of the paths by computing path probability and we say that a path is the rarest if it has the lowest probability. Note that, to compute path probability, one can collect the path constraints using symbolic execution. In this paper, we do not use symbolic execution to collect path constraints. Instead we use a heuristic to compute path probabilities (discussed in section~\ref{section:rare-paths}) that focuses on branch conditions and their selectivity. 

After identifying the rare paths, we guide concolic execution (discussed in section~\ref{section:seed-generation}) to generate inputs that trigger the rare paths. For example,
for path 4 in Table~\ref{tab:intra-inter-paths}, concolic execution generates the input \texttt{DOC}. We provide this input as the initial seed to AFL and we find that AFL can generate the sequences \texttt{DOC} and \texttt{ATT} within 40 minutes (on average) whereas AFL with a random seed cannot generate these sequence in an hour. 

We also collect paths using inter-procedural control flow analysis (paths from 20 to 43 in Table~\ref{tab:intra-inter-paths}). Using our rare path analysis, we identify path 35 as the rarest one. Guiding concolic execution using path 35, the input generated is \text{DOC\textless ATT}. Providing this input as initial seed, fuzzer immediately explores the path covering sequences \texttt{DOC} and \texttt{ATT}.

Using inter-procedural control flow analysis, we can generate the rarest paths in the program. However, paths based on inter-procedural analysis also traverse \texttt{parse\_cmt} which is not necessary to generate the desired sequences \texttt{DOC} and \texttt{ATT} that enable us to explore deeper behaviors. Although, for our small running example, analyzing the procedure \texttt{parse\_cmt} will not waste too much analysis time, for larger real world cases like \texttt{libxml2}, focusing only on inter-procedural paths is likely be costly and can increase the cost of rare path analysis significantly.

To improve the effectiveness of rare path analysis (in order to generate a higher number of rare seeds within a given time budget) we introduce a new kind of control flow path in this paper which we call II-paths (discussed in section~\ref{section:program-paths}). II-paths subsume intra-procedural and inter-procedural control flow paths, and include more paths that combine their characteristics. All the paths in Table~\ref{tab:intra-inter-paths} are II-paths, where paths 1 to 5 are intra-procedural control flow paths, and paths 20 to 43 are inter-procedural control flow paths. Furthermore, paths 6 to 19 are also II-paths. Let us assume that, given a time budget, we can generate the paths from 1 to 20 only. Then, we will identify II-path 13 as the rarest one and concolic execution can generate the input \texttt{DOCATT}. As a result, we will able to generate an input containing sequences \texttt{DOC} and \texttt{ATT} while analyzing a relatively small number of paths.


\section{Program Paths}
\label{section:program-paths}

First step in rare-path guided fuzzing is identification of rare paths. The paths we identify are control flow paths that are generated by traversing control flow graphs of programs. 


\begin{figure}[t]
\begin{lstlisting}[showstringspaces=false,frame=lines,language=C, basicstyle=\scriptsize\selectfont\ttfamily, numberstyle=\scriptsize,mathescape=true]
char *CUR;
#define CMP3( s, c1, c2, c3 ) \
  ( ((unsigned char *) s)[ 0 ] == c1 && \
    ((unsigned char *) s)[ 1 ] == c2 && \
    ((unsigned char *) s)[ 2 ] == c3 )
int main(int argc, char **argv) {
  CUR = argv[1];
  if (CMP3(CUR, 'D', 'O', 'C')) {
    CUR = CUR + 3;
    parse_cmt();
    if(parse_att())
      /* go deeper */
  }
  return 0;
}
void parse_cmt() {
  if(*CUR == '<' $\mid\mid$ *CUR == '>')
    CUR++;
}
int parse_att() {
  if (CMP3(CUR, 'A', 'T', 'T'))
    return 1;
  return 0;
}
\end{lstlisting}
\caption{A code fragment based on the libxml file parser.c.}
\label{fig:running-example}
\end{figure}

\subsection{Control Flow Graphs}

We define the control flow graph (CFG)~\cite{cfg} $G_{\textit{pr}}$ for a procedure \textit{pr} as follows:

\begin{definition}
A control flow graph for a procedure \textit{pr} is a directed graph $G_{\textit{pr}} = (V,E)$ where each vertex $v \in V$ represents a basic block of \textit{pr}, and each directed edge $e \in E: v \rightarrow v'$ represents a possible flow of control from vertex $v$ to vertex $v' \in E$ . 
Control flow graph $G_{\textit{pr}}$ has a unique entry vertex $\textit{entry-pr} \in V$ with no incoming edges and a unique exit vertex $\textit{exit-pr} \in V$ with no outgoing edges.
Furthermore, for each procedure call statement $C$ to a procedure $\textit{pr'}$, $G_{\textit{pr}}$ contains a call vertex $\textit{call-pr'}_C \in V$ and a return-site vertex $\textit{return-pr'}_C \in V$, and an edge $\textit{call-pr'}_C \rightarrow \textit{return-pr'}_C \in E$ that represents the procedure call.

\end{definition}

\begin{figure}
\begin{scriptsize}
    \centering
    \includegraphics[scale=0.12]{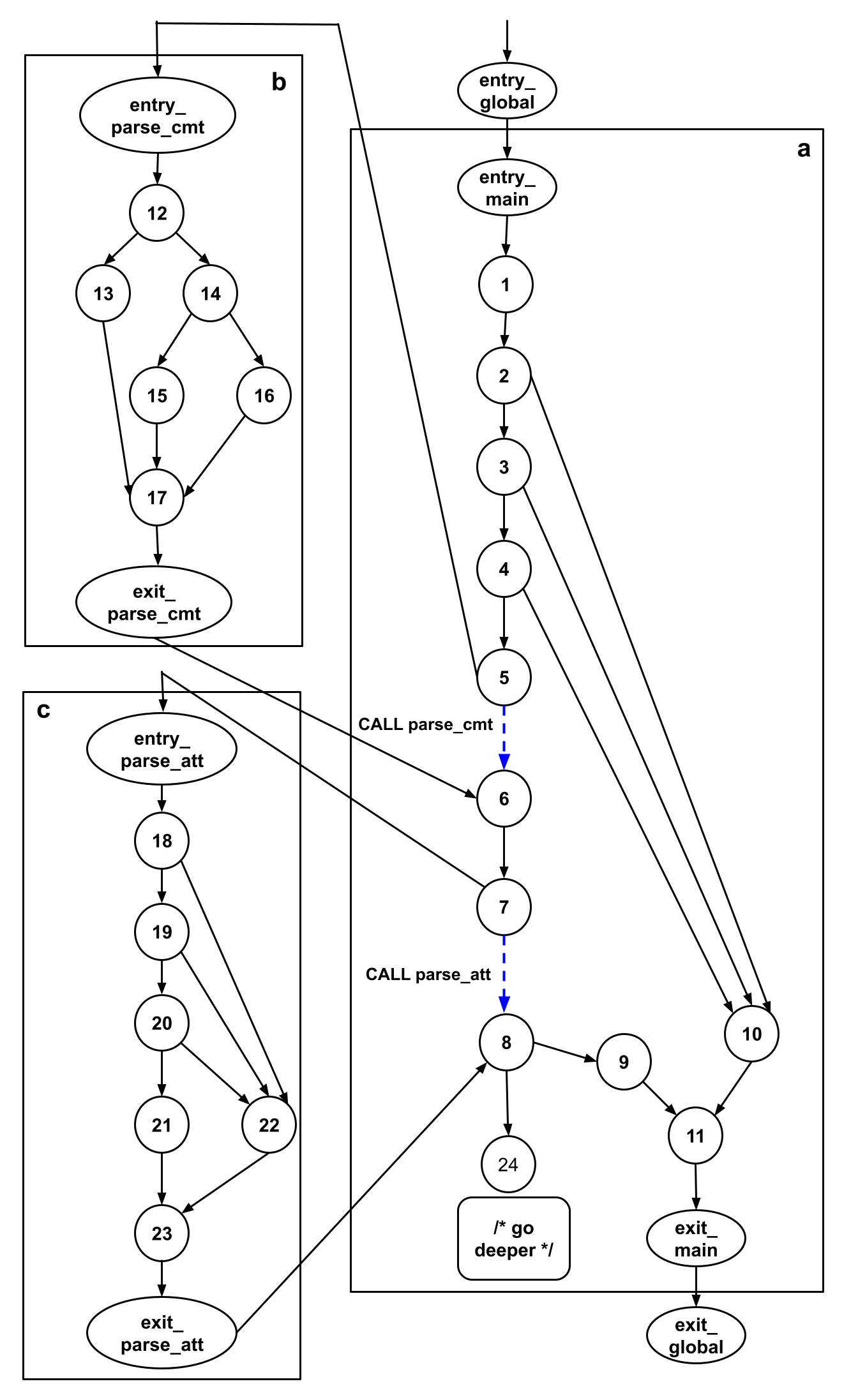}
    \caption{Control flow graphs for the code from Fig.~\ref{fig:running-example}.}
    \label{fig:interprocedural-cfg}
\end{scriptsize}
\end{figure}



Fig.~\ref{fig:interprocedural-cfg} shows the control flow graphs for procedures shown in Fig.~\ref{fig:running-example} in boxes a (\texttt{main}), b (\texttt{parse\_cmt}) and c (\texttt{parse\_att}).




An inter-procedural control flow graph represents control flow of the whole program by combining the control flow graphs of all procedures of the program. 
\begin{definition}
An Inter-Procedural Control Flow Graph (IP-CFG) for a program $P$, $G^+_P = (V,E)$, contains the vertices and edges of the CFGs of all procedures in $P$,
except the edges that correspond to procedure calls. Instead, 
for each procedure call statement $C$ to a procedure $\textit{pr}$ in $P$, $G^+_P$ contains an edge from the call vertex to the entry vertex of the called procedure, $\textit{call-pr}_C \rightarrow \textit{entry-pr} \in E$, and an edge from the exit vertex of the called procedure to the return-site vertex for the call, $\textit{exit-pr} \rightarrow \textit{return-pr}_C \in E$, but it does not contain an edge between the call vertex and the return-site vertex, $\textit{call-pr}_C \rightarrow \textit{return-pr}_C \not \in E$.
$G^+_P$  also contains a vertex $\textit{entry-global} \in V$ with no incoming edges (entry point of the program) and another vertex $\textit{exit-global} \in V$ with no outgoing edges (exit point of the program), and connects them to the main procedure of the program $P$ with edges  
$\textit{entry-global} \rightarrow  \textit{entry-main} \in E$ and $\textit{exit-main} \rightarrow \textit{exit-global} \in E$.
\end{definition}

Fig.~\ref{fig:interprocedural-cfg} shows the IP-CFG for the code from Fig.~\ref{fig:running-example} (the dashed edges are not part of the IP-CFG). For the call to procedure \texttt{parse\_cmt}, there are two edges. One edge from call vertex $5$ (which corresponds to $\textit{call-pr}_C$)
to \textit{entry-parse\_cmt} and one from \textit{exit-parse\_cmt} to return-site vertex $6$
(which corresponds to $\textit{return-pr}_C$).

\subsection{Control Flow Paths}



We define intra- and inter-procedural control flow paths as follows:
\begin{definition}
    Given a control flow graph $G_{\textit{pr}} = (V,E)$ for a procedure \textit{pr}, an intra-procedural control flow path (intra-path) is a sequence of vertices $(v_1, v_2, v_3, \dots, v_n)$ where $\forall i, v_i \in V, v_i \rightarrow v_{i+1} \in E$, $v_1 = \textit{entry\_pr}$ and
    $v_n = \textit{exit\_pr}$.
\end{definition}

\begin{definition}
    Given an inter-procedural control flow graph $G^+_P = (V,E)$ for a program $P$, an inter-procedural control flow path (inter-path) 
    is a sequence of vertices $(v_1, v_2, v_3, \dots, v_n)$ where $\forall i, v_i \in V, v_i \rightarrow v_{i+1} \in E$, $v_1 = \textit{entry-global}$ and
    $v_n = \textit{exit-global}$.
\end{definition}


Paths 1 to 5 in Table~\ref{tab:intra-inter-paths} correspond to all the intra-paths for the CFG of procedure \textit{main}, and paths 20 to 43 in Table~\ref{tab:intra-inter-paths} are all the inter-paths for the IP-CFG of the whole program based on the control flow graphs shown in Fig.~\ref{fig:interprocedural-cfg} for our running example. To save space, we only show the vertices with numeric labels in Table~\ref{tab:intra-inter-paths}.

\subsection{Intra-Inter Control Flow Paths (II-Paths)}


We introduce a new type of control flow paths by combining both intra-paths and inter-paths. We call these paths intra-inter control flow paths (II-paths). Intuitively, for each procedure call, inter-paths have to choose a path inside the called procedure's CFG. On the other hand, intra-paths do not explore the CFGs of the called procedures. When visiting a procedure call statement, II-paths have the option to either behave like intra-paths (i.e., do not explore the CFG of the called procedure), or behave like inter-paths (i.e., explore the CFG of the called procedure).


In order to formally define II-paths we add back an extra edge to the IP-CFG between the call vertex $\textit{call-pr}_C$ and return-site vertex $\textit{return-pr}_C$ for each call statement $C$ (as we had for the intra-procedural control flow graphs in Definition 1). We call the resulting control flow graph Extended Inter-Procedural Control Flow Graph (EIP-CFG):
\begin{definition}
The Extended  Inter-Procedural Control Flow Graph (EIP-CFG) for program $P$,
denoted as $G^\star_P = (V',E')$, is defined using the IP-CFG $G^+_P = (V,E)$ of the program $P$,
where $V'=V$ and $E \subseteq E'$.
The only edges that are in $E'$ and not in $E$ are: For each procedure call statement $C$, a single edge between the call vertex $\textit{call-pr}_C$ and the return-site vertex $\textit{return-pr}_C$ is included in $E'$, i.e., $\textit{call-pr}_C \rightarrow \textit{return-pr}_C \in E'$ whereas
$\textit{call-pr}_C \rightarrow  \textit{return-pr}_C \not \in E$.
\end{definition}


Fig.~\ref{fig:interprocedural-cfg} shows the EIP-CFG for our running example from Fig.~\ref{fig:running-example} where the dashed edges are also part of the EIP-CFG.
In the EIP-CFG,
there are two edges from each call vertex $\textit{call-pr}_C$ for a procedure call: 1) to the entry vertex of called procedure $pr$ $\textit{entry\_pr}$, i.e., edge 
$\textit{call-pr}_C \rightarrow \textit{entry-pr}$
and 2) to the return-site vertex $\textit{return-pr}_C$, i.e., edge 
$\textit{call-pr}_C \rightarrow \textit{return-pr}_C$.
For example, in Fig.~\ref{fig:interprocedural-cfg},
the call vertex $5$ has two outgoing edges corresponding to these two cases 1)
$5 \rightarrow \textit{entry-parse\_cmt}$ and 
2) $5 \rightarrow 6$. 
As a result, whenever a call vertex is reached, there are two different paths to explore: 1) path taken via edge 
$\textit{call-pr}_C \rightarrow \textit{entry-pr}$ which is similar to inter-paths, and
2) path taken via edge $\textit{call-pr}_C \rightarrow \textit{return-pr}_C$ which is similar to intra-paths. 
Intuitively, every time a procedure call vertex is reached, II-paths can choose between considering or ignoring the control flow inside the called procedure. Whereas, intra-paths never explore the control flow of called procedures, and inter-paths always have to explore the control flow of the called procedures. 


We define II-paths as follows:
\begin{definition}
    Given an EIP-CFG $G^\star_P = (V,E)$ for a program $P$, an intra-inter control flow path (II-path) 
    is a sequence of vertices $(v_1, v_2, v_3, \dots, v_n)$ where $\forall i, v_i \in V, v_i \rightarrow v_{i+1} \in E$, $v_1 = \textit{entry-global}$ and
    $v_n = \textit{exit-global}$.
\end{definition}

Again, let us consider the paths (listed in Table~\ref{tab:intra-inter-paths}) of the EIP-CFG shown in Fig.~\ref{fig:interprocedural-cfg} for our running example from Fig.~\ref{fig:running-example}.
As we noted before, paths 1 to 5 in Table~\ref{tab:intra-inter-paths} are all the intra-paths for procedure \textit{main}, and paths 20 to 43 in Table~\ref{tab:intra-inter-paths} are all the inter-paths for the program. Note that, based on the II-paths definition these paths are also II-paths.
Furthermore, using the II-paths definition, in addition to II-paths from 1 to 5 and from 20 to 43, we now have additional II-paths from 6 to 19 where paths from 6 to 13 that ignore the control flow inside procedure \texttt{parse\_cmt} but consider the control flow inside procedure \texttt{parse\_att} and paths from 14 to 19 that ignore the control flow inside procedure \texttt{parse\_att} but consider the control flow inside procedure \texttt{parse\_cmt}. 


\begin{table}[hbt!]
\captionof{table}{II-paths for the extended inter-procedural control flow graph shown in  Fig.~\ref{fig:interprocedural-cfg}.}
\centering
\label{tab:intra-inter-paths}
\scalebox{0.55}{
\begin{tabular}{l|l|l}
    \toprule
     & Path & Probability\\
     \midrule
    1 & $1\rightarrow2\rightarrow10\rightarrow11$ & \num{9.96e-01}\\
\hline
2 & $1\rightarrow2\rightarrow3\rightarrow10\rightarrow11$ & \num{3.98e-03}\\
\hline
3 & $1\rightarrow2\rightarrow3\rightarrow4\rightarrow10\rightarrow11$ & \num{1.59e-05}\\
\hline
4 & $1\rightarrow2\rightarrow3\rightarrow4\rightarrow5\rightarrow6\rightarrow7\rightarrow8\rightarrow9\rightarrow11$ & \num{3.20e-08}\\
\hline
5 & $1\rightarrow2\rightarrow3\rightarrow4\rightarrow5\rightarrow6\rightarrow7\rightarrow8\rightarrow24$ & \num{3.20e-08}\\
\hline
6 & $1\rightarrow2\rightarrow3\rightarrow4\rightarrow5\rightarrow6\rightarrow7\rightarrow18\rightarrow22\rightarrow23\rightarrow8\rightarrow9\rightarrow11$ & \num{3.19e-08}\\
\hline
7 & $1\rightarrow2\rightarrow3\rightarrow4\rightarrow5\rightarrow6\rightarrow7\rightarrow18\rightarrow22\rightarrow23\rightarrow8\rightarrow24$ & \num{3.19e-08}\\
\hline
8 & $1\rightarrow2\rightarrow3\rightarrow4\rightarrow5\rightarrow6\rightarrow7\rightarrow18\rightarrow19\rightarrow22\rightarrow23\rightarrow8\rightarrow9\rightarrow11$ & \num{1.27e-10}\\
\hline
9 & $1\rightarrow2\rightarrow3\rightarrow4\rightarrow5\rightarrow6\rightarrow7\rightarrow18\rightarrow19\rightarrow22\rightarrow23\rightarrow8\rightarrow24$ & \num{1.27e-10}\\
\hline
10 & $1\rightarrow2\rightarrow3\rightarrow4\rightarrow5\rightarrow6\rightarrow7\rightarrow18\rightarrow19\rightarrow20\rightarrow22\rightarrow23\rightarrow8\rightarrow9\rightarrow11$ & \num{5.10e-13}\\
\hline
11 & $1\rightarrow2\rightarrow3\rightarrow4\rightarrow5\rightarrow6\rightarrow7\rightarrow18\rightarrow19\rightarrow20\rightarrow22\rightarrow23\rightarrow8\rightarrow24$ & \num{5.10e-13}\\
\hline
12 & $1\rightarrow2\rightarrow3\rightarrow4\rightarrow5\rightarrow6\rightarrow7\rightarrow18\rightarrow19\rightarrow20\rightarrow21\rightarrow23\rightarrow8\rightarrow9\rightarrow11$ & \num{2.05e-15}\\
\hline
13 & $1\rightarrow2\rightarrow3\rightarrow4\rightarrow5\rightarrow6\rightarrow7\rightarrow18\rightarrow19\rightarrow20\rightarrow21\rightarrow23\rightarrow8\rightarrow24$ & \num{2.05e-15}\\
\hline
14 & $1\rightarrow2\rightarrow3\rightarrow4\rightarrow5\rightarrow12\rightarrow13\rightarrow17\rightarrow6\rightarrow7\rightarrow8\rightarrow9\rightarrow11$ & \num{1.28e-10}\\
\hline
15 & $1\rightarrow2\rightarrow3\rightarrow4\rightarrow5\rightarrow12\rightarrow13\rightarrow17\rightarrow6\rightarrow7\rightarrow8\rightarrow24$ & \num{1.28e-10}\\
\hline
16 & $1\rightarrow2\rightarrow3\rightarrow4\rightarrow5\rightarrow12\rightarrow14\rightarrow15\rightarrow17\rightarrow6\rightarrow7\rightarrow8\rightarrow9\rightarrow11$ & \num{1.27e-10}\\
\hline
17 & $1\rightarrow2\rightarrow3\rightarrow4\rightarrow5\rightarrow12\rightarrow14\rightarrow15\rightarrow17\rightarrow6\rightarrow7\rightarrow8\rightarrow24$ & \num{1.27e-10}\\
\hline
18 & $1\rightarrow2\rightarrow3\rightarrow4\rightarrow5\rightarrow12\rightarrow14\rightarrow16\rightarrow17\rightarrow6\rightarrow7\rightarrow8\rightarrow9\rightarrow11$ & \num{3.17e-08}\\
\hline
19 & $1\rightarrow2\rightarrow3\rightarrow4\rightarrow5\rightarrow12\rightarrow14\rightarrow16\rightarrow17\rightarrow6\rightarrow7\rightarrow8\rightarrow24$ & \num{3.17e-08}\\
\hline
20 & $1\rightarrow2\rightarrow3\rightarrow4\rightarrow5\rightarrow12\rightarrow13\rightarrow17\rightarrow6\rightarrow7\rightarrow18\rightarrow22\rightarrow23\rightarrow8\rightarrow9\rightarrow11$ & \num{1.27e-10}\\
\hline
21 & $1\rightarrow2\rightarrow3\rightarrow4\rightarrow5\rightarrow12\rightarrow13\rightarrow17\rightarrow6\rightarrow7\rightarrow18\rightarrow22\rightarrow23\rightarrow8\rightarrow24$ & \num{1.27e-10}\\
\hline
22 & $1\rightarrow2\rightarrow3\rightarrow4\rightarrow5\rightarrow12\rightarrow13\rightarrow17\rightarrow6\rightarrow7\rightarrow18\rightarrow19\rightarrow22\rightarrow23\rightarrow8\rightarrow9\rightarrow11$ & \num{5.10e-13}\\
\hline
23 & $1\rightarrow2\rightarrow3\rightarrow4\rightarrow5\rightarrow12\rightarrow13\rightarrow17\rightarrow6\rightarrow7\rightarrow18\rightarrow19\rightarrow22\rightarrow23\rightarrow8\rightarrow24$ & \num{5.10e-13}\\
\hline
24 & $1\rightarrow2\rightarrow3\rightarrow4\rightarrow5\rightarrow12\rightarrow13\rightarrow17\rightarrow6\rightarrow7\rightarrow18\rightarrow19\rightarrow20\rightarrow22\rightarrow23\rightarrow8\rightarrow9\rightarrow11$ & \num{2.04e-15}\\
\hline
25 & $1\rightarrow2\rightarrow3\rightarrow4\rightarrow5\rightarrow12\rightarrow13\rightarrow17\rightarrow6\rightarrow7\rightarrow18\rightarrow19\rightarrow20\rightarrow22\rightarrow23\rightarrow8\rightarrow24$ & \num{2.04e-15}\\
\hline
26 & $1\rightarrow2\rightarrow3\rightarrow4\rightarrow5\rightarrow12\rightarrow13\rightarrow17\rightarrow6\rightarrow7\rightarrow18\rightarrow19\rightarrow20\rightarrow21\rightarrow23\rightarrow8\rightarrow9\rightarrow11$ & \num{8.19e-18}\\
\hline
27 & $1\rightarrow2\rightarrow3\rightarrow4\rightarrow5\rightarrow12\rightarrow13\rightarrow17\rightarrow6\rightarrow7\rightarrow18\rightarrow19\rightarrow20\rightarrow21\rightarrow23\rightarrow8\rightarrow24$ & \num{8.19e-18}\\
\hline
28 & $1\rightarrow2\rightarrow3\rightarrow4\rightarrow5\rightarrow12\rightarrow14\rightarrow15\rightarrow17\rightarrow6\rightarrow7\rightarrow18\rightarrow22\rightarrow23\rightarrow8\rightarrow9\rightarrow11$ & \num{1.27e-10}\\
\hline
29 & $1\rightarrow2\rightarrow3\rightarrow4\rightarrow5\rightarrow12\rightarrow14\rightarrow15\rightarrow17\rightarrow6\rightarrow7\rightarrow18\rightarrow22\rightarrow23\rightarrow8\rightarrow24$ & \num{1.27e-10}\\
\hline
30 & $1\rightarrow2\rightarrow3\rightarrow4\rightarrow5\rightarrow12\rightarrow14\rightarrow15\rightarrow17\rightarrow6\rightarrow7\rightarrow18\rightarrow19\rightarrow22\rightarrow23\rightarrow8\rightarrow9\rightarrow11$ & \num{5.08e-13}\\
\hline
31 & $1\rightarrow2\rightarrow3\rightarrow4\rightarrow5\rightarrow12\rightarrow14\rightarrow15\rightarrow17\rightarrow6\rightarrow7\rightarrow18\rightarrow19\rightarrow22\rightarrow23\rightarrow8\rightarrow24$ & \num{5.08e-13}\\
\hline
32 & $1\rightarrow2\rightarrow3\rightarrow4\rightarrow5\rightarrow12\rightarrow14\rightarrow15\rightarrow17\rightarrow6\rightarrow7\rightarrow18\rightarrow19\rightarrow20\rightarrow22\rightarrow23\rightarrow8\rightarrow9\rightarrow11$ & \num{2.03e-15}\\
\hline
33 & $1\rightarrow2\rightarrow3\rightarrow4\rightarrow5\rightarrow12\rightarrow14\rightarrow15\rightarrow17\rightarrow6\rightarrow7\rightarrow18\rightarrow19\rightarrow20\rightarrow22\rightarrow23\rightarrow8\rightarrow24$ & \num{2.03e-15}\\
\hline
34 & $1\rightarrow2\rightarrow3\rightarrow4\rightarrow5\rightarrow12\rightarrow14\rightarrow15\rightarrow17\rightarrow6\rightarrow7\rightarrow18\rightarrow19\rightarrow20\rightarrow21\rightarrow23\rightarrow8\rightarrow9\rightarrow11$ & \num{8.16e-18}\\
\hline
35 & $1\rightarrow2\rightarrow3\rightarrow4\rightarrow5\rightarrow12\rightarrow14\rightarrow15\rightarrow17\rightarrow6\rightarrow7\rightarrow18\rightarrow19\rightarrow20\rightarrow21\rightarrow23\rightarrow8\rightarrow24$ & \num{8.16e-18}\\
\hline
36 & $1\rightarrow2\rightarrow3\rightarrow4\rightarrow5\rightarrow12\rightarrow14\rightarrow16\rightarrow17\rightarrow6\rightarrow7\rightarrow18\rightarrow22\rightarrow23\rightarrow8\rightarrow9\rightarrow11$ & \num{3.16e-08}\\
\hline
37 & $1\rightarrow2\rightarrow3\rightarrow4\rightarrow5\rightarrow12\rightarrow14\rightarrow16\rightarrow17\rightarrow6\rightarrow7\rightarrow18\rightarrow22\rightarrow23\rightarrow8\rightarrow24$ & \num{3.16e-08}\\
\hline
38 & $1\rightarrow2\rightarrow3\rightarrow4\rightarrow5\rightarrow12\rightarrow14\rightarrow16\rightarrow17\rightarrow6\rightarrow7\rightarrow18\rightarrow19\rightarrow22\rightarrow23\rightarrow8\rightarrow9\rightarrow11$ & \num{1.26e-10}\\
\hline
39 & $1\rightarrow2\rightarrow3\rightarrow4\rightarrow5\rightarrow12\rightarrow14\rightarrow16\rightarrow17\rightarrow6\rightarrow7\rightarrow18\rightarrow19\rightarrow22\rightarrow23\rightarrow8\rightarrow24$ & \num{1.26e-10}\\
\hline
40 & $1\rightarrow2\rightarrow3\rightarrow4\rightarrow5\rightarrow12\rightarrow14\rightarrow16\rightarrow17\rightarrow6\rightarrow7\rightarrow18\rightarrow19\rightarrow20\rightarrow22\rightarrow23\rightarrow8\rightarrow9\rightarrow11$ & \num{5.06e-13}\\
\hline
41 & $1\rightarrow2\rightarrow3\rightarrow4\rightarrow5\rightarrow12\rightarrow14\rightarrow16\rightarrow17\rightarrow6\rightarrow7\rightarrow18\rightarrow19\rightarrow20\rightarrow22\rightarrow23\rightarrow8\rightarrow24$ & \num{5.06e-13}\\
\hline
42 & $1\rightarrow2\rightarrow3\rightarrow4\rightarrow5\rightarrow12\rightarrow14\rightarrow16\rightarrow17\rightarrow6\rightarrow7\rightarrow18\rightarrow19\rightarrow20\rightarrow21\rightarrow23\rightarrow8\rightarrow9\rightarrow11$ & \num{2.03e-15}\\
\hline
43 & $1\rightarrow2\rightarrow3\rightarrow4\rightarrow5\rightarrow12\rightarrow14\rightarrow16\rightarrow17\rightarrow6\rightarrow7\rightarrow18\rightarrow19\rightarrow20\rightarrow21\rightarrow23\rightarrow8\rightarrow24$ & \num{2.03e-15}\\
\bottomrule
  \end{tabular}
}

\end{table}


\section{Identifying Rare Paths}
\label{section:rare-paths}


\begin{figure}
 \begin{scriptsize}
     \centering
     \includegraphics[scale=0.12]{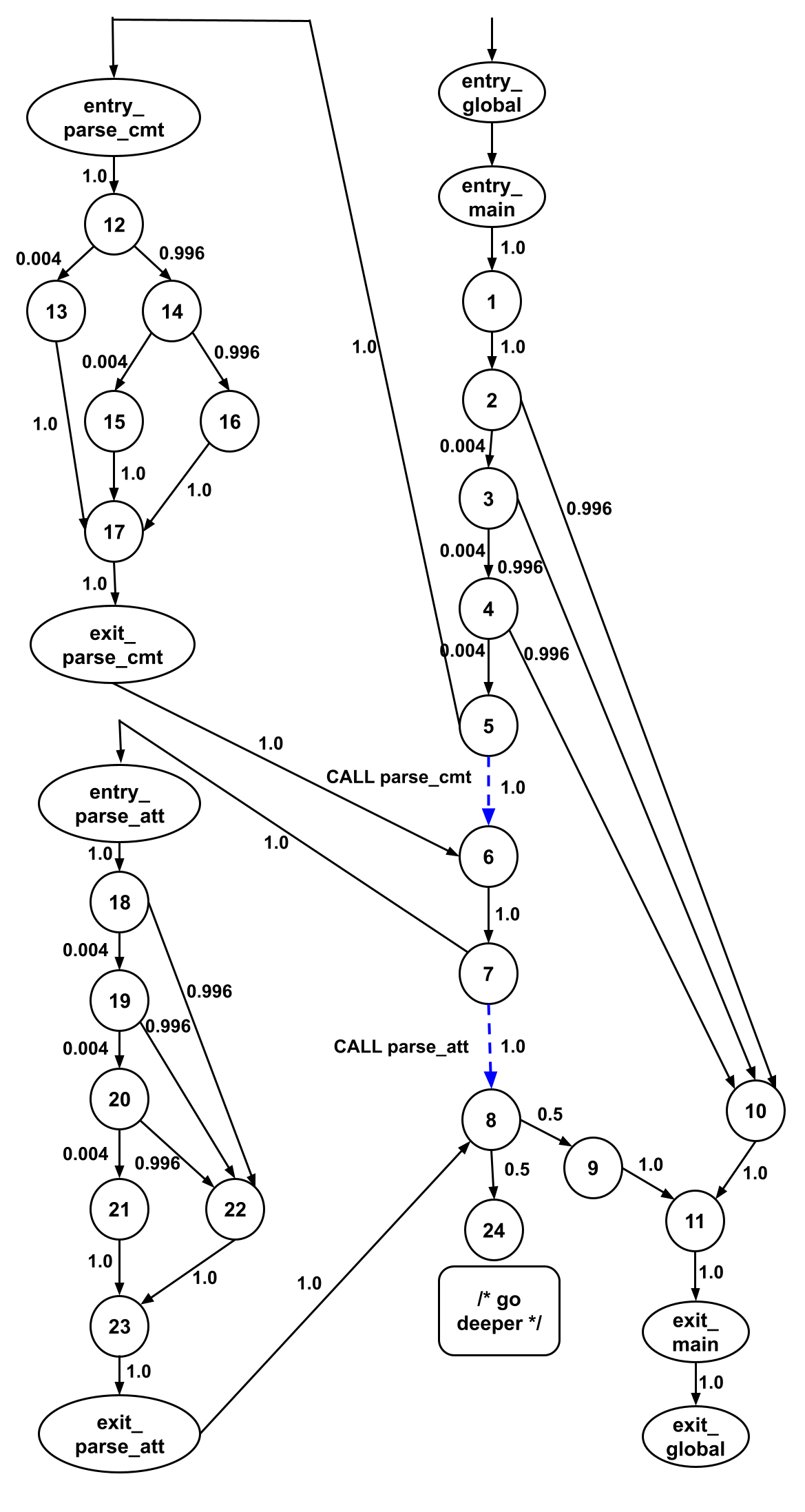}
 \caption{Probabilistic inter-procedural control flow graph for the code from
 Fig.~\ref{fig:running-example}.} 
%
\label{fig:prob-interprocedural-cfg}
\end{scriptsize}
\end{figure}

In this section, we describe construction of a probabilistic control flow graph to compute path probabilities. Then, we identify the rare paths based on path probabilities.

\subsection{Path Probability} 
Given a program $P$, let $i$ denote the input for the program, and $I$ denote the domain of inputs (i.e., $i \in I$). Given a path  $t$ in program $P$, the goal of path probability analysis is to determine how likely it is to execute the path $t$. We do this by determining the likelihood of picking inputs that result in the execution of path $t$. In order to determine the likelihood of picking such inputs, we compute the probability of picking such inputs if the inputs are chosen randomly. We define  $\mathcal{P}(P, t)$ as:

\begin{definition}
$\mathcal{P}(P, t)$ denotes the probability of executing the path $t$ of program $P$ where the input $i$ of the program $P$ is randomly selected from the input domain $I$.
\end{definition}

To compute path probability, we assume that inputs are uniformly distributed. However, one can extend our technique for path probability computation by integrating usage profile~\cite{statistical-se}, used in other probabilistic analysis techniques and support any input distribution.

Path probabilities can be computed using quantitative extensions of symbolic execution such as probabilistic and statistical symbolic execution~\cite{probabilistic-se,statistical-se}. However, these symbolic execution based techniques have a high computation complexity and poor scalibility due to the cost of path constraint solving and model counting over an exponentially increasing number of paths. Recently, a new heuristic-based technique has been proposed for probabilistic reachability analysis~\cite{preach}, which reduces the complexity of probabilistic  analysis using a concept called branch selectivity. In this paper, we focus on 
computing path probabilities using branch selectivity instead of computing reachability probabilities of program statements using a Discrete-Time Markov Chain model as in~\cite{preach}. 


\subsection{Probabilistic Control Flow Graph}
To compute path probabilities, we introduce the concept of the probabilistic control flow graph (Prob-CFG). Prob-CFG $PG^\star_P$ for a program $P$ is constructed using the extended inter-procedural control flow graph (EIP-CFG) $G^\star_P$ for $P$. We define the probabilistic control flow graph $PG^\star_P$ as follows:
\begin{definition}
Given a program $P$ and its EIP-CFG $G^\star_P = (V, E)$, the probabilistic control flow graph $PG^\star_P$ for program $P$ is defined as $PG^\star_P = (V, E, F)$ where the set of vertices and edges for  $PG^\star_P$ are same as the set of vertices and edges of $G^\star_P$, and $F$ is a function 
$F: E \rightarrow [ 0, 1 ]$ that assigns a probability score to each edge in $E$.


\end{definition}

As we describe below, we use dependency analysis and branch selectivity to compute probability scores of the edges in 
probabilistic control flow graphs. 

\paragraph*{Dependency Analysis} A branch condition in the program is input dependent if the evaluation of the branch condition depends on the value of the program input. Given a program and input(s) to the program, we use static dependency analysis to identify the input dependent branch vertices in the control flow graph. Static dependency analysis over-approximates the set of input-dependent branch vertices. As a result, the path probability we compute is an estimation of the actual path probability. Anyway, we use branch selectivity, a heuristic to estimate path probability. 

\paragraph*{Branch Selectivity} To compute the probability for each edge in the control flow graph, we use branch selectivity. We use the definition of branch selectivity $\mathcal{S}(b)$ as in~\cite{preach}:

\begin{definition}
Given a branch condition $b$, let $D_b$ denote the Cartesian product of the domains of the variables that appear in $b$, and let $T_b \subseteq D_b$ denote
the set of values for which branch condition $b$
evaluates to true. Let $|D_b|$
and $|T_b|$ denote the number of 
elements in these sets, respectively.
Then, $\mathcal{S}(b) = \frac{|T_b|}{|D_b|}$ and $0\leq \mathcal{S}(b)\leq 1$.
\end{definition}

We compute $|T_b|$ using a model counting constraint solver. 
Branch selectivity gets closer to 0 as the number of values that satisfy the branch condition decreases and gets closer to 1 as the number of values that satisfy the branch condition increases. 

We define the probability score function $F$ for the probabilistic control flow graph $PG_P = (V, E, F)$ using the combination of dependency analysis and branch selectivity as follows:
\begin{itemize}
\item If there is only one edge starting from a vertex $v$ to $u$, then the probability of the edge $e: v \rightarrow u$ is 1, i.e, $F(e) =1$. 
\item If $v$ is a vertex with branch condition $b$, there are two edges from source vertex $v$: $e_1: v \rightarrow u_1$ and $e_2: v \rightarrow u_2$, where $e_1$ is the true evaluation and $e_2$ is the false evaluation of branch condition $b$:
\begin{itemize}
\item If branch condition $b$ is dependent on program input, then probability of edge $e_1$ is the branch selectivity, $\mathcal{S}(b) = \frac{|T_b|}{|D_b|}$ and the probability of edge $e_2$ is $1 - \mathcal{S}(b)$,
i.e., $F(e_1) = \mathcal{S}(b)$ and $F(e_2) = 1 - \mathcal{S}(b)$.
\item If branch condition $b$ is not dependent on program input, then probability of both edges $e_1$ and $e_2$ is 1,
i.e., $F(e_1) = F(e_2) = 1$. 
\end{itemize}
\item Probabilities of edges that have a call vertex as their source $e_1 : \textit{call-pr}_C \rightarrow \textit{entry-pr}$ and $e_2 : \textit{call-pr}_C \rightarrow \textit{return-pr}_C$ are 1, i.e., 
$F(e_1) = F(e_2) = 1$.
\end{itemize}

By adding probabilities to all the edges in a control flow graph, we transform it to a probabilistic control flow graph. Consider the EIP-CFG in Fig.~\ref{fig:interprocedural-cfg}. Each branch vertex is associated with a branch condition. For example, vertex 2 is associated with branch condition \texttt{CUR[0] = D}. 
We consider that the inputs are uniformly distributed and domain for each character in a string has 256 values. Branch selectivity $\mathcal{S}$ for the branch condition at vertex 2 is ${1\over256} \equiv 0.004$. Hence, probability for the edges $2\rightarrow3$ is $0.004$ and probability for the edge $2\rightarrow10$ is $1 - 0.004 = 0.996$. 
We add all the edge probabilities to the EIP-CFG $G^\star_P$ in Fig.~\ref{fig:interprocedural-cfg} and construct the probabilistic EIP-CFG $PG^\star_P$, shown in Fig.~\ref{fig:prob-interprocedural-cfg}.


Once we construct the probabilistic control flow graph $PG^\star_P$, we can compute path probabilities as follows:
\begin{definition}
Given a control flow path $t$ for program $P$ which corresponds to a sequence of vertices $\{v_1,v_2,v_3,\dots,v_n\}$ in the probabilistic control flow graph $PG^\star_P = (V, E, F)$, then path probability $\mathcal{P}(P,t)$ for path $t$ is computed as
\[ \mathcal{P}(P,t) = \prod_{i=1}^{n-1} F(v_i, v_{i+1}) \]
\end{definition}
Path probabilities computed for the II-paths for our running example using the probabilistic control flow graph in Fig.~\ref{fig:prob-interprocedural-cfg} are shown in Table~\ref{tab:intra-inter-paths}.

\subsection{Rare Paths}
We call a program path a rare path if it is unlikely to be executed when the program input is randomly chosen. 
Since there may be an unbounded number of paths in a program, given a depth bound $b$, we identify the set of $k$ rare paths among all paths with length less than or equal to $b$.
\begin{definition}
Given a number $k$, a program $P$ and a bound $b$, the set of $k$ rare paths $R = \{t_1, t_2, t_3, ..., t_k\}$
are the set of paths with length less than or equal to $b$ and with lowest probabilities, 
i.e., $ \forall t, t \in R \Rightarrow |t| \leq b \wedge
\forall  t',  |t'| \leq b \wedge t' \not \in R  \Rightarrow \mathcal{P}(P,t) \leq \mathcal{P}(P,t')$. 
\end{definition}




Traversing through the probabilistic control flow graph in Fig.~\ref{fig:interprocedural-cfg} we generate 43 II-paths and compute corresponding path probabilities as shown in Table~\ref{tab:intra-inter-paths}. Now, if we sort these paths in an ascending order based on the path probability and pick the set of rare paths $R$ for $k = 3$, we identify paths 34, 35 and 26 as the paths in the rare path set $R$. A fuzzer that randomly generates inputs would be very unlikely to explore these rare paths.




\section{Input Generation for Rare Paths}
\label{section:seed-generation}

The analysis we described above results in the set $R$ of $k$ rare paths in the program. However, it does not identify $k$ inputs that can trigger these rare paths in the program. The input generation process we describe in this section identifies inputs to trigger the rare paths in the set $R$. 


In order to generate the set of rare inputs $I_R$ for the set of rare paths $R$ we guide concolic execution using each rare path $t_R \in R$ and generate input $i_R$ for each $t_R$ (if path $t_R$ is a feasible execution path). We add all these inputs to the set of rare inputs $I_R$. 

Note that, the rare paths we compute are based on an estimation of path probability and some of the rare paths might not be feasible. But, concolic execution captures the original program execution semantics. Hence, if a rare path is not feasible, it will be eliminated in the input generation step using concolic execution.

We use path-guided concolic execution to collect path constraints for a rare path. We then use a SMT solver to solve the path constraints and generate the input that can be fed to the program to execute the rare path. 
We provide two different algorithms for path-guided concolic execution for input generation: 1) Inter-path guided concolic execution, 2) II-path guided concolic execution.


\subsection{Inter-path guided Concolic Execution (IP-GCE)}
For inter-path guided concolic execution (\textsc{IP-GCE}), we run the program on a concrete random input and generate the corresponding inter-path $t_C$. In order to generate input for the rare path $t_R$, we compare all branches for $t_C$ and $t_R$ in the same order. If there is a mismatch between any of the branches, we negate the branch and solve it to check feasibility of the path negating the branch. If the path is feasible, we solve the path constraint and generate the new input. We then execute the program using the new input and update $t_C$ by the inter-path generated by the new input. The process continues as long as there are branches left to compare both in $t_C$ and $t_R$ or there are no branches that can lead to a feasible path. At the end of the process, the input is the input that will either take path $t_R$ or take a path that is close to the rare path $t_R$ if $t_R$ is not feasible. 

Algorithm~\ref{proc:inter-procedural-path-guided-concolic-execution} shows the process of guiding concolic execution using rare inter-path. \textsc{Execute} executes the program $P$ first on a random input and returns the corresponding execution path $t_C$. The algorithm looks for the first vertex where $t_C$ and $t_R$ differ (all paths start with the same vertex).
$\textsc{NegatedPath}(t_C, \textit{index})$ generates a path constraint corresponding to the path $t_C$ where the branch condition between the vertex $\textit{index} -1$ and $\textit{index}$ is negated and all the branches before $\textit{index}-1$ remain the same. \textsc{IsFeasible} checks the feasibility of a given path constraint and 
\textsc{Solve} generates an input value satisfying the given path constraint.


\begin{algorithm}
\caption{IP-GCE($P, t_R$)\\
 Takes a program $P$ and an inter-procedural path $t_R$ as input and generates an input for $P$ to execute the path $t_R$}
\label{proc:inter-procedural-path-guided-concolic-execution}
\begin{footnotesize}
\begin{algorithmic}[1]
\State $\textit{input} \leftarrow \textsc{Random}()$
\State $t_C \leftarrow \textsc{Execute}(P, \textit{input})$
\State $\textit{index} \leftarrow 2$
\While{$\textit{index} < \textsc{Len}(t_C) \land \textit{index} < \textsc{Len}(t_R)$}
  \If{$t_C(\textit{index}) \neq t_R(\textit{index})$}
     \State $\textit{path\_cond} \leftarrow \textsc{NegatedPath}(t_C, \textit{index})$
    \If{$\textsc{IsFeasible}(\textit{path\_cond})$} 
       \State $\textit{input} \leftarrow \textsc{Solve}(\textit{path\_cond})$ 
       \State $t_C \leftarrow \textsc{Execute}(P, \textit{input})$
    \Else
      \State \Return $\textit{input}$
    \EndIf
  \EndIf
  \State $\textit{index} \leftarrow \textit{index + 1}$
\EndWhile
\State \Return $\textit{input}$
\end{algorithmic}
\end{footnotesize}
\end{algorithm} 


\subsection{II-Path Guided Concolic Execution (IIP-GCE)}
In this section we discuss II-Path guided concolic execution which can also handle intra-paths since intra-paths are also II-paths. IP-GCE algorithm we discussed in the previous section uses branch matching and branch negation for mismatched branches, but this approach is not sufficient for guiding the concolic execution to explore the rare II-paths since  II-paths are not guaranteed to represent complete execution path of a program.  

Similar to the \textit{IP-GCE} algorithm, in the \textit{IIP-GCE} algorithm (Algorithm~\ref{proc:intra-ii-procedural-path-guided-concolic-execution}),  we first run the program on a concrete random input and collect the execution path $t_C$. 
Note that, there may be branches in $t_C$ that are in a procedure that is not explored in the input II-path $t_R$. In such situations, we compare the inputs that trigger both the branch and its negation, and see which one creates an execution path that overlaps more with $t_R$ (i.e., increases the number of vertices that are common in both), and then we pick the branch which results in higher overlap with $t_R$.


Lines 1-5 in Algorithm~\ref{proc:intra-ii-procedural-path-guided-concolic-execution} generate the initial concrete path $t_C$ with a random input, and calculate the initial overlap between $t_C$ and $t_R$ using the function $\textsc{Overlap}$. 

The while loop in lines 6-19 iterates over the nodes in $t_C$. It looks for branch nodes in $t_C$ that differ from the corresponding branch node in $t_R$. The function $\textsc{Differ}$ returns true under two conditions: 1) there is a branch in $t_R$ that corresponds to complement of $t_C(\textit{index})$ (i.e., $t_R$ and $t_C$ take different branches for the same branch statement), or 2) there is no
branch in $t_R$ that corresponds to the branch $t_C(\textit{index})$
(this branch node in $t_C$ corresponds to a branch in a procedure that was
not explored in $t_R$).
In both of these cases we negate the branch condition at $t_C(\textit{index})$ and see if we can improve the overlap between $t_C$ and $t_R$, and update the input and $t_C$ if the overlap can be improved. Note that, if the overlap cannot be improved, then the input is restored to the previous input in lines 17-18. 

Algorithm~\ref{proc:intra-ii-procedural-path-guided-concolic-execution} makes a single pass on the branches in $t_C$ without backtracking and therefore it is not guaranteed to find an execution that maximizes the overlap between final $t_C$ and $t_R$. Looking for maximum overlap would require a search on all execution paths, resulting in path explosion that we have to avoid for scalability.

\begin{algorithm}[t]
\caption{IIP-GCE($P, t_R$)\\
Takes a program $P$ and an II-path $t_R$ as input and generates an input for $P$ to execute a path that has high overlap with $t_R$}
\label{proc:intra-ii-procedural-path-guided-concolic-execution}
\begin{footnotesize}
\begin{algorithmic}[1]
\State $\textit{input} \leftarrow \textsc{Random}()$
\State $t_C \leftarrow \textsc{Execute}(P, \textit{input})$
\State $\textit{max\_overlap} \leftarrow \textsc{Overlap}(t_C, t_R)$
\State $\textit{max\_input} \leftarrow \textit{input}$ 
\State $\textit{index} \leftarrow 1$
\While{$\textit{index} < \textsc{Len}(t_C)$}
 \If{$\textsc{IsBranch}(t_C(\textit{index})) \wedge \textsc{Differ}(t_C(\textit{index}), t_R)$}
 \State $\textit{path\_cond} \leftarrow \textsc{NegatedPath}(t_C, \textit{index})$
 \If{$\textsc{IsFeasible}(\textit{path\_cond})$} 
 \State $\textit{input} \leftarrow \textsc{Solve}(\textit{path\_cond})$ 
 \State $t_C \leftarrow \textsc{Execute}(P, \textit{input})$
 \State $\textit{overlap} \leftarrow \textsc{Overlap}(t_C, t_R)$
 \If{$\textit{overlap} > \textit{max\_overlap}$}
  \State $\textit{max\_overlap} \leftarrow \textit{overlap}$
  \State $\textit{max\_input} \leftarrow \textit{input}$
  \Else
  \State $\textit{input} \leftarrow \textit{max\_input}$
  \State $t_C \leftarrow \textsc{Execute}(P, \textit{input})$
  \EndIf
  \EndIf
 \EndIf
 \State $\textit{index} \leftarrow \textit{index} + 1$
\EndWhile
\State \Return $\textit{input}$
\end{algorithmic}
\end{footnotesize}
\end{algorithm}

For the running example, guiding concolic execution using path 35, input generated is \texttt{DOC\textless ATT}.
whereas guiding concolic execution using path 34, we find out that path 34 is infeasible. Path 34 is infeasible as path up to vertex 8 in path 34, $parse\_att$ function returns 1 and then returning back to the main function it should take the path following edge $8\rightarrow24$ whereas it takes edge $8\rightarrow9$. Hence, path-guided concolic execution algorithms we provide does not only generate inputs but also checks feasibility of the rare paths. Even though our techniques for identifying rare paths in the program is a heuristic approach, infeasible rare paths will be always filtered out in the input generation phase. The inputs we generate are always valid inputs and they help fuzzer in exploring rare program paths. 

\section{Implementation}
\label{sec:implmentation}
We implemented our techniques for rare path analysis and path-guided concolic execution analyzing programs written in the C programming language.

We extract branch conditions and control flow graph for a program
using the concolic execution tool CREST~\cite{crest} and underlying program transformation tool CIL~\cite{cil}. In order to collect branch conditions from the program, we modified the OCaml code in CIL. We transform the branch conditions in the input program to constraints in the SMT-LIB format. To model count the branch constraints, we use the Automata-based Model Counter (ABC)~\cite{ABB15}.

To identify input dependent branches in the program, we perform dependency analysis using CodeQL~\cite{codeql} code analysis engine. 
To implement dependency analysis, we used the \textsc{Access} module of CodeQL that provides classes for modeling accesses including variable accesses, enum constant accesses and function accesses.

After extracting the control flow graph and model counting the input dependent branches, we transform the control flow graph to a probabilistic control flow graph. We wrote python scripts to traverse the probabilistic control flow graph and collect intra-, inter- and II-paths.

We guide concolic execution tool CREST~\cite{crest} using the rare paths we collect from our control flow analysis. We wrote algorithms \textsc{IP-GCE} and \textsc{IIIP-GCE} in C on top of the existing concolic search strategies in CREST.

We use existing coverage-guided fuzzers AFL++~\cite{afl++} and FairFuzz~\cite{fairfuzz} as it is. We contacted the authors of DigFuzz~\cite{digfuzz} but an implementation was not publicly available. We implemented DigFuzz using AFL++ and QSym~\cite{qsym}. To collect edge coverage we use \texttt{afl-showmap} as used in~\cite{wu2022evaluating}. 


\section{Experimental Evaluation}
\label{sec:experiments}

To evaluate our techniques for rare path-guided fuzzing we experiment on a set of benchmarks (programs with many restrictive branch conditions)
that have already been used in experimental evaluation of existing fuzzing techniques. \textit{inih} (parser for .ini configuration file), \textit{tinyC} (parser for tiny C codes with if-else, while, do-while structures), \textit{cJSON} (parser for JSON files) have been used for evaluating parser-directed fuzzing~\cite{parser-directed-fuzzing}. We also add \textit{calculator}~\cite{calculator} (a command-line calculator, supporting standard mathematical operations and a set of function), more complex in terms of restrictive branch conditions. We also experiment on two well known libraries for parsing xslt and xml files, \textit{libxslt} and \textit{libxml2} respectively. \textit{libxslt} has been used in~\cite{skyfire} and \textit{libxml2} has been used to evaluate many coverage guided fuzzing techniques~\cite{afl,afl-fast,fairfuzz}.


In our experimental evaluation we focused on the following research questions: \\
\textbf{RQ1.} Can rare path analysis generate inputs that AFL++ can not? \\
\textbf{RQ2.} Can we improve fuzzing effectiveness using the seed set we generate from our rare path analysis? \\
\textbf{RQ3.} Can we improve rare path analysis effectiveness using II-paths?

\subsection{Experimental Setup}
We run our experiments on a virtual box equipped with an Intel Core i7-8750H CPU at 2.20GHz and 16 GB of RAM running Ubuntu Linux 18.04.3 LTS. We use docker for AFL++~\cite{afl++-docker} to run all the fuzzing experiments. We run each fuzzing task with a random seed set for 24 hours. We set the upper limit for our rare path guidance technique (branch selectivity computation, rare path identification and seed generation) to 6 hours (25\% of the total time) and use the remaining 18 hours (of 24 hour total time) fuzzing with the seed set generated by our analysis. We set path depth limit to 60 for our rare path analysis. After collecting the rare paths, we provide all the inputs from the feasible rare paths (filtered by path-guided concolic execution) to the fuzzer as the seed set.

\subsection{Experimental Results}
\subsubsection{RQ1: Effectiveness of rare path analysis to generate rare inputs}
To show the effectiveness of our rare path analysis, we run our analysis maximum for 6 hours and AFL++ for 24 hours on each of these benchmarks. Our experimental results show that we can generate inputs in 6 hours which AFL++ cannot generate in 24 hours by mutating inputs. Our results in detail are as follows.

\textbf{tinyC.} 
We generate inputs containing \textit{if-else} structure from our rare path analysis. AFL++ can generate \textit{if} structure by mutating inputs but cannot generate the \textit{if-else} structure. 

\textbf{inih.} 
Each ini file has section names inside  an opening bracket, [  and a closing bracket, ] and key value pairs separated by either a colon (:) or an equal sign (=). From our rare path analysis we can find these rare input structures within a minute. But, AFL++ can also generate these inputs within couple of minutes as the input structure is trivial. So, for \textit{inih}, we cannot generate any new inputs. 

\textbf{calculator.} 
We generate inputs containing keywords such as \textit{arcsin}, \textit{arccos} and \textit{arctan} with our rare path analysis. Even after running AFL++ for 24 hours, AFL++ cannot generate these keywords. 


\textbf{cJSON.} 
AFL++ can generate inputs containing basic JSON structure with left and right braces, colon and quotations. But, using our rare path analysis, we can generate inputs containing keywords such as \textit{false}, \textit{true} and \textit{null} that AFL++ is unable to generate.

\textbf{libxslt.} 
To explore deeper paths in the program xslt files need to contain keywords like \texttt{stylesheet}, \texttt{transform}, \texttt{attribute-set}, \texttt{preserve-space}, \texttt{decimal-format} etc. As a random seed, we provide XSLT file containing opening and closing tag for \texttt{stylesheet} to AFL++. However, running AFL++ for 24 hours, it cannot generate inputs containing any other keywords. Our rare path analysis can generate inputs containing keywords: \texttt{attribute-set}, \texttt{preserve-space} and \texttt{decimal-format}. 

\textbf{libxml2.} 
Similar to \textit{libxslt}, to explore deeper paths in \textit{libxml2}, a xml file needs to contain keywords like \texttt{DOCTYPE}, \texttt{ATTLIST}, \texttt{ENTITY}, \texttt{NOTATION} etc. Running AFL++ for 24 hours, it can generate inputs containing structures like \texttt{DOCTYPE} and \texttt{ATTLIST}. Our rare path analysis can generate inputs containing not only \texttt{DOCTYPE} and \texttt{ATTLIST} but also \texttt{ENTITY} and \texttt{NOTATION}.

Overall, we see that for 5 out 6 benchmarks, within 6 hours (25\% of the time allocated to AFL++), our rare path analysis can generate inputs that AFL++ cannot generate in 24 hours based on input mutation. 

\subsubsection{RQ2: Effectiveness of rare path analysis to improve fuzzing effectiveness}
Our answer to RQ1 already shows that the rare path analysis can generate inputs that AFL++ cannot. Now, to answer RQ2, we present experimental results evaluating the ability of rare path analysis in improving fuzzing effectiveness in terms of coverage. 

\begin{figure}
\begin{scriptsize}
    \centering
    \includegraphics[scale=0.3]{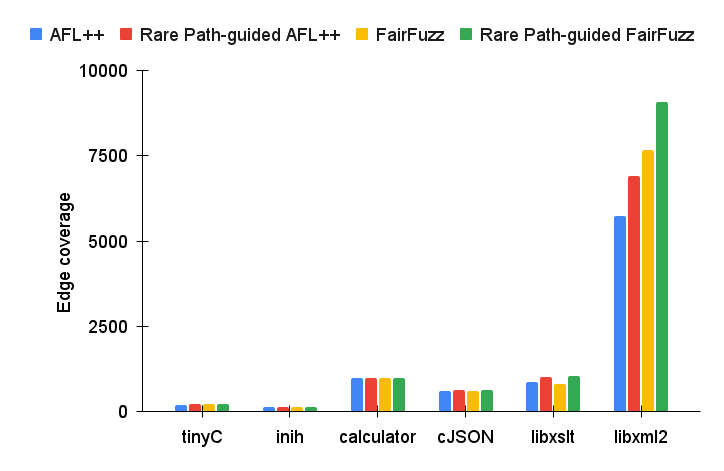}
    \caption{Coverage comparison between AFL++, rare-path guided AFL++, FairFuzz and rare-path guided FairFuzz}
    \label{fig:coverage-comparison}
\end{scriptsize}
\end{figure}

Our experimental results show that (as shown in Fig.~\ref{fig:coverage-comparison} and Table~\ref{tab:percentage-coverage-improvement}) we get coverage improvement over AFL++ for 5 out of 6 of the benchmarks. We do not get a lot of improvement for \textit{calculator} (1.13\%) since, even though we can generate rare inputs, there are no deeper functionalities to execute after passing through the rare branches. We generate inputs containing functions: \textit{arcsin}, \textit{arccos}, \textit{arctan} using our rare path analysis. And with these additional inputs, AFL++ can mutate and generate 3 more rare inputs: \textit{asin}, \textit{acos}, \textit{atan}. However, there are not many functionalities to explore and code to cover after these rare branches. AFL++ with the rare path based seed set can cover only 13 additional edges (1.33\% improvement). For \textit{tinyC} and \textit{cJSON}, we see improvement of 6.47\% (13 additional edges) and 4.19\% (25 additional edges), respectively. For \textit{libxslt}, our rare path guidance helps AFL++ to cover 162 additional edges (18.86\% coverage improvement). For \textit{libxml2}, we achieve the maximum amount of coverage improvement of 1170 additional edges (20.35\%). This indicates that for larger programs if restrictive branches in the program can be passed, fuzzers can explore deeper functionalities and achieve significantly more code coverage, and our rare path analysis can guide the fuzzers to pass the restrictive branches in the program. 

Next, we experimentally evaluate our rare path analysis using FairFuzz~\cite{fairfuzz} using the same setup that we used for AFL++. For 5 out of 6 cases, we see improvement, 0.51\% for \textit{calculator} 0.94\% for \textit{tinyC}, 4.14\% for \textit{cJSON} and 18.29\% for \textit{libxml2} (shown in Fig.~\ref{fig:coverage-comparison} and Table~\ref{tab:percentage-coverage-improvement}). The results are similar to AFL++, for larger programs, FairFuzz can explore more deeper functionalities and achieve more code coverage. For \textit{libxslt}, FairFuzz without any guidance can cover 800 edges whereas with guidance it can cover 1055 edges (31.86\% coverage improvement). For \textit{libxml2}, FairFuzz without inputs from our analysis can cover 7681 edges, whereas with guidance from rare path analysis, it can cover 9086 edges (18.29\% of coverage improvement).

Moreover, for \textit{cJSON}, \textit{libxslt} and \textit{libxml2}, our rare path analysis can generate inputs that FairFuzz cannot. This indicates that FairFuzz (which uses branch hit counts to identify rare branches) can not pass some rare branches.
However, we can identify and generate inputs for these rare branches. Rare path guided FairFuzz performs best in our experimental evaluation (1.33\%, 5.36\%, 7.46\%, 22.82\% and 58.00\% more coverage than AFL++ for \textit{calculator}, \textit{cJSON}, \textit{tinyC}, \textit{libxslt} and \textit{libxml2} respectively).



\begin{table}[hbt!]
\captionof{table}{Percentages of coverage improvement for rare path-guided fuzzing over AFL++, FairFuzz}
\centering
\label{tab:percentage-coverage-improvement}
\begin{footnotesize}
  \begin{tabular}{l|r|r|r}
    \toprule
  Benchmarks & Number of lines & \multicolumn{2}{c}{\% coverage improvement over}\\
  & & \ \ \ \ \ \ \ AFL++  & FairFuzz\\
  \midrule
    tinyC &	190	& 6.47\% & 0.94\%\\
    inih &	243	& 0.00\% & 0.00\%\\
    calculator & 1312 &	1.33\% & 0.51\%\\
    cJSON &	3845 & 4.19\% & 4.14\%\\
    libxslt & 33371 & 18.86\% & 31.86\%\\
    libxml2	& 186116 & 20.35\% & 18.29\%\\
    \bottomrule
  \end{tabular}
\end{footnotesize}
\end{table}


Lastly, we evaluate our rare path analysis on top of hybrid fuzzing technique, DigFuzz~\cite{digfuzz}. DigFuzz~\cite{digfuzz} identifies the hardest paths to explore for AFL using the samples collected using AFL and then uses symbolic execution tool angr~\cite{angr} to solve constraints for the hardest paths. However, DigFuzz is not publicly available. We contacted the authors of DigFuzz but could not get access to the implementation. Hence, we implement the technique in DigFuzz using AFL++ and QSym~\cite{qsym}. 
In our evaluation we use an unoptimized binary for fuzzing (to associate branch flip in concolic execution with hitcount collected in fuzzing which is necessary for the implementation of the DigFuzz technique). 
We conduct experiments on the 3 larger benchmarks, \textit{cJSON}, \textit{libxslt} and \textit{libxml2}. Results from our experimental evaluation (Table~\ref{tab:digfuzz}) show that rare path guided DigFuzz achieves better coverage compared to DigFuzz, 66.86\% improvement for \textit{cJSON}, 2.18\% improvement for \textit{libxslt} and 30.22\% improvement for \textit{libxml2}.

There are multiple reasons behind DigFuzz not being able to achieve better coverage compared to AFL++ and FairFuzz: 1) building the execution tree takes hours for larger programs like \textit{libxml2} as the tree grows exponentially over time, 2) concolic execution fails to generate inputs for a lot of paths and hence generates very few inputs to guide AFL and 3) DigFuzz attempts to solve branches that are not dependent on the inputs rather used for sanity check of the program. These findings are aligned to the findings of DigFuzz for larger programs~\cite{digfuzz}. However, our experiments on DigFuzz still demonstrate that rare path guided analysis improves the effectiveness of DigFuzz like it improves AFL++ and FairFuzz.

\begin{table}[hbt!]
\captionof{table}{Percentages of coverage improvement for rare path-guided fuzzing over DigFuzz}
\centering
\label{tab:digfuzz}
\begin{footnotesize}
  \begin{tabular}{l|r|r|r}
    \toprule
  Benchmarks & DigFuzz & \makecell{Rare Path-guided\\DigFuzz} & \makecell{\% coverage\\improvement}\\
  \midrule
    cJSON &	344 & 574 & 66.86\%\\
    libxslt & 719 & 735 & 2.18\%\\
    libxml2	& 3297 & 4270 & 30.22\%\\
    \bottomrule
  \end{tabular}
\end{footnotesize}
\end{table}

\subsubsection{RQ3: Effectiveness of II-path to improve efficiency of rare path analysis}
\begin{figure*}
\centering
\begin{scriptsize}
\begin{subfigure}[b]{0.33\textwidth}
    \includegraphics[scale=0.33]{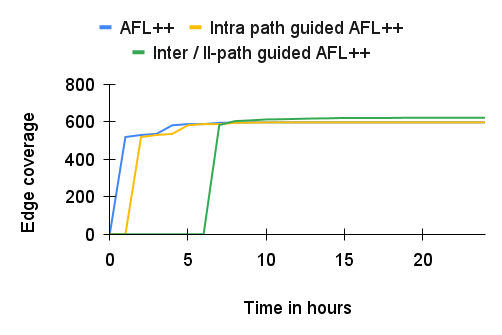}
    \caption{cJSON}
    \label{fig:cjson-iipath-coverage-improvement}
\end{subfigure}
\hfill
\begin{subfigure}[b]{0.33\textwidth}
    \centering
    \includegraphics[scale=0.33]{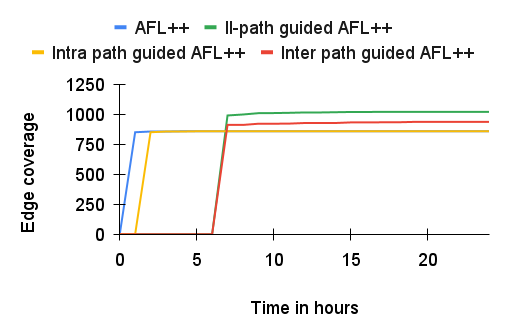}
    \caption{libxslt}
    \label{fig:libxslt-iipath-coverage-improvement}
\end{subfigure}
\hfill
\begin{subfigure}[b]{0.33\textwidth}
    \centering
    \includegraphics[scale=0.33]{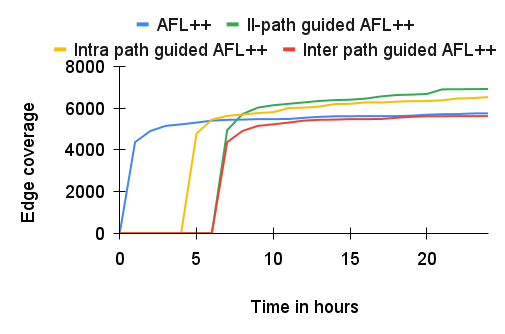}
    \caption{libxml2}
    \label{fig:libxml2-iipath-coverage-improvement}
\end{subfigure}
\end{scriptsize}
    \caption{Coverage improvement comparison between different types of path-guided fuzzing. II-paths can generate more number of rare inputs compared to both intra and inter paths within a given amount of time and hence highest edge coverage is achieved by II-path guided fuzzing.}
    \label{fig:all-iipath-coverage-improvement}
\end{figure*}

To answer RQ2, we guide fuzzers using our rare path analysis based on intra-paths, inter-paths and II-paths. Our claim is that II-paths based analysis can generate more rare inputs compared to either intra-paths or inter-paths or both. Our experimental results for \textit{cJSON}, \textit{libxslt} and \textit{libxml2} is shown in (Fig.~\ref{fig:all-iipath-coverage-improvement}) respectively. 

Using intra paths for \textit{cJSON}, we do not see any improvements as we cannot generate any inputs. However, using inter paths we can generate inputs and see improvements (4.19\%). Using II-paths we can also generate the same inputs and see same amount of coverage improvement.

For \textit{libxslt}, using intra paths, we do not see any improvement as it can can not generate new inputs. Using inter paths, we see coverage improvement (9.08\%) as new inputs are generated containing keywords \texttt{preserve-space} and \texttt{decimal-format}. However, using II-paths, we see the highest improvement (17.93\%) as inputs containing keyword (\texttt{attribute-set}) is also generated.

For \textit{libxml2}, using inter paths, we do not see any improvements rather coverage is reduced as we waste 25\% of the fuzzing time analyzing the paths. Inter paths can not find any rare inputs as it goes deep inside each and every procedure. Some of these procedures being analyzed for rare paths do not contain any complex program checks and due to exponential increase in the number of paths, it wastes time and cannot analyze significant procedures that contains complex program checks. Hence, the identified rare paths based on inter-paths are not actually rare paths for \textit{libxml2} and guiding concolic execution using these rare paths does not generate rare inputs that can improve coverage performance.

Identifying rare paths based on intra paths for \textit{libxml2} can generate an input containing the specific values: \texttt{DOCTYPE} and hence, we see coverage improvement.
It can generate \texttt{DOCTYPE} as the branch conditions comparing to this specific value were inside the initial starting procedure. 
II-paths can generate inputs containing specific values: \texttt{DOCTYPE}, \texttt{ATTLIST}, \texttt{ENTITY} and \texttt{NOTATION}. These inputs help to achieve better coverage not only compared to AFL++ (20.35\%) but also compared to both intra rare path (5.99\%) and inter rare path (23.18\%) analysis.

\section{Related Work}
\label{sec:relwork}
\paragraph*{Mutation-based coverage guided Fuzzers} 
AFL~\cite{afl} is a well-known mutation-based coverage guided fuzzer. AFL++~\cite{afl++} is the latest version of AFL with more speed, better mutation techniques, better instrumentation and support from custom modules. 
In this work, we use default version of AFL++ which uses power schedule of AFLFast~\cite{afl-fast}. There are a lot of mutation-based coverage guided fuzzers focusing on advanced mutation strategies. MOPT~\cite{mopt} focuses on mutation scheduling by providing different probabilities to the mutation operators. LAF-INTEL~\cite{laf-intel} focuses on bypassing hard multibyte comparisons, by splitting them into multiple single-byte comparison. REDQUEEN~\cite{redqueen} focuses on bypassing Input-To-State (I2S) defined comparisons. 
Steelix\cite{steelix} performs static analysis and extra instrumentation to produce inputs satisfying multi-byte comparisons.  VUzzer~\cite{vuzzer} identifies input positions used in the comparison and immediate values using a Markov Chain model and decides which parts of the program should be targeted. FairFuzz~\cite{fairfuzz} identifies the rare branches in the program based on the hitcounts of branches. If a rare branch is identified by FairFuzz, it applies input mutation masking.

In this paper, we focus on identifying rare program paths. We neither use a fuzzer to identify rare paths, nor modify mutation strategies inside the fuzzer. We show that we can improve the effectiveness of state of the art fuzzers without making any changes to the internals of fuzzers.

\paragraph*{Symbolic execution guided fuzzers}
Hybrid fuzzing techniques~\cite{driller,digfuzz, deepfuzzer} use symbolic execution and constraint solvers to generate inputs to pass complex checks in the program. Driller~\cite{driller} uses selected symbolic execution when fuzzer can not cover new branches for a long period of time. DigFuzz~\cite{digfuzz} uses the fuzzer itself to statistically identify hardest paths for the fuzzer to explore and then uses symbolic execution to solve path constraints for the hardest paths. 
DeepFuzzer~\cite{deepfuzzer} uses lightweight symbolic execution to pass initial complex checks and then it relies on seed selection and mutation techniques. In this work, we do not use the path samples from fuzzer to identify rare paths, rather we statically analyze programs. Moreover, we do not symbolically execute the whole program and instead guide symbolic execution using the rare paths we identify to generate inputs.

\paragraph*{Grammar-based Fuzzers} Grammar-based fuzzing techniques generate well-formed inputs based on a user provided grammar~\cite{godefroid2008grammar,yoo2016grammar}. These fuzzing techniques mutate inputs using the derivative rules in the grammar. As a result, the mutated input is also guaranteed to be well-formed~\cite{liang2018fuzzing}. Grammar-based fuzzers are very effective to fuzz programs that are heavily dependent on structured inputs~\cite{godefroid2008grammar, yan2013structurized}. However, grammar-based fuzzers require application specific knowledge of the program under test. There are several fuzzers~\cite{lzfuzz, finding, code-fragments, parser-directed-fuzzing, skyfire} focusing specifically on structured inputs such as fuzzing network protocols, compilers, parser for json, xml, xslt files etc. Compared to grammar based fuzzing, the technique we provide is general, it does not require any knowledge about the program under test and it is fully automated. We neither need to provide an input grammar, nor feed inputs to the parser~\cite{parser-directed-fuzzing, yan2013structurized} or collect large data samples~\cite{skyfire} like techniques that specialize on structured inputs.

\paragraph*{Seed generation for fuzzers}
There are fuzzing techniques that focus on seed selection and seed prioritization to improve fuzzing efficiency~\cite{spotfuzz, slf, seed-selection}. SpotFuzz~\cite{spotfuzz} identifies invalid execution and time consuming edges as hot spots based on hitcounts of different inputs on the edges 
SLF~\cite{slf} is a technique which focuses on valid seed input generation. It 
performs sophisticated input mutation to get through the validity checks.~\cite{seed-selection} systematically investigates
and evaluates the affect of seed selection on fuzzer’s ability to find bugs and demonstrates that fuzzing outcomes vary depending on the initial seeds used. In this work, we also demonstrate that rare inputs as initial seeds bootstraps the fuzzer. However, 
we focus on generating seeds that can execute rare paths.

\paragraph*{Static program analysis for fuzzing}
A large number of fuzzing techniques~\cite{redqueen, steelix, vuzzer, greyone, parser-directed-fuzzing, taint-fuzz} use static program analysis techniques to guide fuzzers. Most of these techniques use either control flow analysis or taint analysis. In this work, we also use control flow analysis and dependency analysis to identify rare paths. However we introduce a novel technique we call rare path analysis and a new kind of control flow paths (II-paths). Although different, our definition of II-paths is inspired by the control flow directed concolic search techniques provided in~\cite{crest}.

\section{Conclusions}
\label{sec:conclusion}
In this paper, we provide techniques to identify rare program paths that are difficult for a fuzzer to explore generating random inputs. To identify the rare paths, we use lightweight static analysis. We use the identified rare paths to guide a concolic execution tool to generate inputs that can execute these rare paths. Finally, we provide these inputs as the initial seed set to the fuzzer. From our experimental evaluation on a set of benchmarks, having lots of restrictive branch conditions, we find that we can generate inputs that a fuzzer cannot generate mutating inputs. These inputs from our analysis also guide the fuzzer to achieve better coverage compared to an initial random seed. To speed up or rare path analysis, we also introduced a new type of control flow paths (II-paths) in this paper.

\bibliographystyle{plain}
\bibliography{biblio}

\end{document}